\documentclass[aps,twocolumn,pra,superscriptaddress,preprintnumbers,showpacs,tightenlines]{revtex4}
\usepackage{amssymb}
\usepackage{epsfig,graphicx,times}

\begin{document}

\title{Geometric phases in dressed state quantum computation}
\author{Lian-Ao Wu}
\affiliation{Ikerbasque, Basque Foundation for Science, 48011 Bilbao, Spain}
\affiliation{Department of Theoretical Physics and History of Science, The Basque Country
University (EHU/UPV), PO Box 644, 48080 Bilbao, Spain}
\author{C. Allen Bishop}
\affiliation{Physics Department, Southern Illinois University, Carbondale, Illinois
62901-4401}
\author{Mark S. Byrd}
\affiliation{Physics Department, Southern Illinois University, Carbondale, Illinois
62901-4401}
\affiliation{Computer Science Department, Southern Illinois University, Carbondale,
Illinois 62901}
\date{\today }

\begin{abstract}
Geometric phases arise naturally in a variety of quantum systems with
observable consequences.  They also arise in quantum computations when
dressed states are used in gating operations.  Here we show how they
arise in these gating operations and how one may take advantage of 
the dressed states producing them.  
Specifically, we show that that for a given, but arbitrary
Hamiltonian, and at an arbitrary time $\tau $, there {\sl always
  exists} a set of dressed states such that a given 
gate operation can be performed by the Hamiltonian up to a phase $\phi $.
The phase is a sum of a dynamical phase and a geometric phase. We illustrate
the new phase for several systems.  
\end{abstract}

\maketitle

\emph{Introduction.--} Quantum gates are the building blocks for quantum
computers and are necessary for a vast array of quantum 
mechanical devices. Quantum gates appear in a 
wide variety of forms and range in complexity from 
simple single-qubit rotations to highly sophisticated multi-qubit, or more generally, multi-qudit designs. An ideal gating operation 
will produce a 
coherent unitary evolution of a quantum state in such a way as to realize 
the desired transformation of the system of interest with perfect fidelity. 
Finding a universal set of these operations which can manipulate and 
entangle collections of qubits is one of the most important tasks 
in quantum computing. Indeed, unitary-evolution-based 
quantum computation relies on a 
processor's ability to manipulate superpositions as well as create 
and destroy an entanglement using a sequence of these gates. When a particular gate is required, for whatever 
the reason may be, one typically has a specific Hamiltonian in mind to 
generate the corresponding evolution. In some cases, 
the implementation of a quantum gate can be carried out using the interactions which are 
naturally available to the system, as, for instance, in the case 
of transferring 
information between processors using the always-on Heisenberg 
interactions occurring between sites of a spin chain \cite{Bose}. The situation is usually 
not so simple however, generally one needs to control these 
gating operations using external fields or other mechanisms, such as measurements. In 
order to accomplish a specific computational goal one must first 
identify a physical system which can effectively serve as a collection of
qubits, initiate the system to an appropriate state, and then force the 
system to evolve according to the combined system/external 
Hamiltonian. We will show here that there 
always exists a complete set of states which, when acted upon by 
an arbitrary evolution 
operator, can produce any given gate, up to an additional phase
factor. It is interesting that a part of this phase factor has a
geometrical nature similar to the Aharonov-Anandan (A-A) phase
\cite{AA}.  This shows that proper initialization, which can
experimentally be different from initializing to the ground state, can
be used as a resource for individual quantum gating operations.

Over the years the geometric phase has been a topic of central interest 
in quantum
mechanics \cite{Wilcek}. Unlike the dynamical phase which 
depends on the dynamical evolution of the system, e.g., the speed 
at which the parameterized path is followed or the arbitrary choice 
of gauge, the geometrical phase is completely determined by a closed 
trajectory of the system in the underlying parameter space and does not 
depend on the 
details of the temporal evolution. The geometrical phase is a measurable quantity and can be used to investigate 
the nontrivial geometric properties of the parameter space. 
It can be observed in the interference of two identically prepared 
systems as they develop a relative phase while either system is adiabatically varied. It can also be observed in single quantum systems which 
are prepared in superpositions of eigenstates of the Hamiltonian. In this case each eigenstate can establish a geometric phase as the Hamiltonian is varied, 
and the differences between these phases can be observed by the measurable 
properties of the system.

It was Berry \cite{Berry} who first noticed the fact that a wave function, 
originally in a nondegenerate eigenstate of the initial Hamiltonian,  
will acquire a geometrical phase factor in addition to the familiar 
dynamical phase if the evolution is induced by a Hamiltonian which is varied adiabatically around a closed path in parameter space. 
Later, Aharonov and Anandan \cite{AA} examined a generalization of the Berry 
phase without recourse to adiabaticity. Their work showed how any closed 
path in the projective Hilbert space of state vectors modulo phase factors 
has a geometric phase associated with it regardless of any adiabatic conditions.

Here, we introduce a family of geometric phases which are associated with 
quantum gating operations. As we have just mentioned, we will show 
that it is possible for any Hamiltonian to mimic the action of a 
specified yet arbitrary gate provided that the system is initialized to an 
appropriate state. The geometrical contribution of the phase acquired 
by this quantum evolution can be calculated in a way that is similar 
to the usual prescription given by Aharonov and Anandan. In this 
situation the geometrical phase associated with the gating operation 
can be determined using a dressed Hamiltonian which generates 
a cyclic evolution. 

We will begin our discussion of these results with an example that 
clearly illustrates the nature of the geometrical phase obtained 
by the cyclic evolution of a state of a spin chain. The example 
should help clarify part of the motivation behind this work, as 
it indicates the origin of a geometrical phase acquired by 
the non-cyclic evolution of a state through a gating operation.

\emph{Example of perfect state transfer.--} Let us start with
an example of perfect state transfer\ \cite{Christandl} through a spin chain. 
Consider an $N$%
-site nearest neighbor \textit{XY} model $H=\sum_{j}J_{j}(\sigma
_{j}^{+}\sigma_{j+1}^{-}+\sigma _{j+1}^{+}\sigma _{j}^{-}).$ When we engineer the
tunneling integrals such that $J_{j}=J\sqrt{j(N-j)}/2,$ the Hamiltonian\
becomes $H=JL_{x}$, where $L_{x}$ is the $x$-component of the quasi-angular
momentum operator. 

At time $\tau $ the evolution operator associated with this Hamiltonian 
becomes \cite{wupst}%
\begin{equation}
U(\tau )=\exp (-iJ\tau L_{x})\,.
\end{equation}%
When $\tau =\pi /2J,$ a spin-up state $\left\vert \uparrow
\right\rangle _{1}$ at site 1 is perfectly transferred to $r\left\vert
\uparrow \right\rangle _{N}$ at the last site $N,$ where $r=\exp \left( i\pi
(N-1)/2\right) .$

We first consider the situation when $\tau =\pi /J$ , the system evolves
cyclically, $\left\vert \uparrow \right\rangle _{1}\rightarrow $ $%
r^{2}\left\vert \uparrow \right\rangle _{1}$, except for a phase 
factor $r^{2}$.
If the number of sites $N$ is even, $r^{2}=e^{i\pi (N-1)}=-1,$ which is
apparently a geometric phase factor since it originates from the geometrical
length of the chain and is not related to the dynamics. More precisely, it is
an Aharonov-Anandan (A-A) phase factor. The A-A phase factor exists 
for any cyclic evolution of a quantum system, defined by $\left\vert \psi (\tau )\right\rangle =e^{i\phi
}\left\vert \psi (0)\right\rangle $. 
It has been shown that cyclic evolutions of this type universally 
exist in any quantum system, regardless of the specific Hamiltonian 
which generates the evolution, provided the system begins in an 
appropriate state \cite{Wu94}. Now, we assume that the wave function is driven by the
Schr\"{o}dinger equation 
\begin{equation}
H(t)\left\vert \psi (t)\right\rangle =i\frac{d}{dt}\left\vert \psi
(t)\right\rangle  \label{SE}
\end{equation}
where we let $\hbar = 1$ throughout. Aharonov and Anandan found an 
expression which removed the 
the dynamical part from the total phase $\phi$
\[
\beta =\phi +\int_{0}^{\tau }\left\langle \psi (t)\right\vert H\left\vert
\psi (t)\right\rangle dt=i\int_{0}^{\tau } \left\langle \tilde{\psi}(t)\right\vert
d\left\vert \tilde{\psi}(t)\right\rangle 
\]%
where $\left\vert \tilde{\psi}(t)\right\rangle =e^{-if(t)}\left\vert \psi
(t)\right\rangle $ and $f(\tau )-f(0)=\phi.$ $\beta$ is known as 
the A-A phase and is uniquely defined up to $2\pi n$ for some integer $n$. 
It is a quantity that is independent of both 
$\phi$ and the underlying Hamiltonian $H$. It is easy to check that for 
the example above we have $%
e^{i\beta }=r^{2}$ or $\beta =\pi (N-1)$ \ mod $2\pi $.

The process of perfect state transfer $\left\vert \uparrow
\right\rangle _{1}\rightarrow $ $r\left\vert \uparrow \right\rangle _{N}$ 
does not correspond to a 
cyclic evolution. However, the above analysis implies that the phase 
factor $r$,
which originates similarly to the A-A phase factor $r^{2},$ should possess geometric
attributes as well. We will exhibit the occurance of similar phenomena 
in various
systems after general discussions.

\emph{Geometric phase originating from quantum gates.--} Consider a quantum
unitary gate $\mathcal{G}$, such as a swap gate, or a CNOT gate, and the system
evolution operator $U(t)$. The evolution operator is related to 
the Hamiltonian through the operator Schr\"{o}dinger 
equation $H(t)=i%
\dot{U}(t)U^{\dagger }(t).$ The combined operator $W(t)=\mathcal{G}^{\dagger
}U(t)$ is also unitary since $W^{\dagger }(t)W(t)=U^{\dagger }(t)\mathcal{GG}%
^{\dagger }U(t)=1$. As with any unitary operator, the operator $W(\tau )$, at
time $t=\tau ,$ can be diagonalized and has a complete set of orthonormal
eigenvectors $\left\{ \Psi _{k}(0)\right\} _{\tau }$ and exponential
eigenvalues $\left\{ \exp (i\phi _{k})\right\} _{\tau }$. A vector $\Psi
_{k}(0)$ in the set obeys the eigenequation: 
\begin{equation}
W(\tau )\Psi _{k}(0)=\exp (i\phi _{k})\Psi _{k}(0).  \label{cyclic}
\end{equation}%
Consequently, if the initial state $\left\vert \psi (0)\right\rangle $\
is one of the $\Psi _{k}(0)$'s, the wave function evolves as $\left\vert
\psi (\tau )\right\rangle =\exp (i\phi )\left\vert \psi (0)\right\rangle 
$, driven by an effective Hamiltonian 
\[
\mathcal{H}(t)=i\dot{W}(t)W^{\dagger }(t)=\mathcal{G}^{\dagger }i\dot{U}%
(t)U^{\dagger }(t)\mathcal{G=G}^{\dagger }H(t)\mathcal{G}. 
\]%
In other words, the dynamics driven by the \emph{effective} or \emph{dressed}
Hamiltonian $\mathcal{H}(t)$ is cyclic if the initial state $\left\vert \psi
(0)\right\rangle $\ is one of the $\Psi _{k}(0)$'s. The corresponding $\phi $
is a sum of the dynamic phase of the \emph{effective} Hamiltonian $\mathcal{H%
}(t)$ and the A-A geometric phase $\beta $, which we refer to as the \emph{dressed} A-A
phase. The essence of the A-A phase $\beta $ being \emph{geometric} is
based on the effective dynamics governed by $\mathcal{H}(t)$. However the
effective Hamiltonian may correspond to multiple operators i.e.,  
$\mathcal{H}(t)$ does not change if we replace $W(t) \rightarrow W(t)V$ when 
$V$ is
unitary. 

On the other hand, the wave function $\left\vert \psi (\tau
)\right\rangle $ driven by the \emph{bare} Hamiltonian $H(t)$ satisfies 
\[
\left\vert \psi (\tau )\right\rangle =U(\tau )\left\vert \psi
(0)\right\rangle =\exp (i\phi )\mathcal{G}\left\vert \psi (0)\right\rangle 
\]%
This indicates that for an arbitrary Hamiltonian and at an arbitrary time $%
\tau $, there \emph{universally exists} a set of states such that a given
gate operation can be performed by this Hamiltonian up to a phase $\phi $.
In the case where $\mathcal{G}$ is a unit operator, the\ dressed A-A phase
becomes the normal A-A phase \cite{Wu94}.

In the example above, a perfect state transfer through a spin chain 
requires a gate to exchange
states at the first and last sites. The operator $\mathcal{G}=r^{\ast }\exp
(i\pi L_{x})$ plays this role, where $\mathcal{G}^{2}=1$. It is easy to
check that $\beta =\phi =\pi (N-1)/2$ \ mod $2\pi $ or $e^{i\beta }=r.$ 
Although the evolution is not cyclic, we are able to extract the 
geometrical phase associated with the evolution since it can be 
described by a gating operator which then defines the 
effective Hamiltonian $\mathcal{H}(t)$. $\left\vert \uparrow
\right\rangle _{1}$ is an eigenstate of the operator 
$W(\tau)$ for $\tau =\pi /2J,$ and thus evolves cyclically under 
the action of the dressed Hamiltonian $\mathcal{H}(t)$. This allows 
us to use the expression for the A-A phase, with $H$ replaced 
with $\mathcal{H}$, in order to determine 
the {\it{dressed}} phase associated with this process.

\emph{Universal set of gates in quantum computation and their coexisting
phases.--- }In the case that $\mathcal{G}=\exp (-i\theta _{0}\sigma _{x})$, we
will consider a \textit{non-perturbative} time-dependent Hamiltonian, 
\[
H(t)=\left\{ 
\begin{array}{c}
\varpi \sigma _{z},0<t<\delta \\ 
\omega \sigma _{x},t>\delta%
\end{array}%
\right.  
\]%
where $\sigma _{x}$, $\sigma _{y}$ and $\sigma _{z}$ are the Pauli matrices.
Here, the evolution operator is given by

\[
U(t)=\left\{ 
\begin{array}{c}
\exp (-i\varpi t\sigma _{z}),0<t<\delta \\ 
\exp (-i\omega (t-\delta )\sigma _{x})\exp (-i\varpi \delta \sigma
_{z}),t>\delta%
\end{array}%
\right. , 
\]%
We can control the time such that $\omega (\tau -\delta )=\theta _{0}$ at
time $t=\tau $ and $W(\tau )=\exp (-i\varpi \delta \sigma _{z}).$ The
eigenstates and eigenvalues are $\Psi _{\uparrow ,\downarrow }(0)=$ $%
\left\vert \uparrow ,\downarrow \right\rangle $ and $\phi _{\uparrow
}=-\varpi \delta ;$ $\phi _{\downarrow }=\varpi \delta .$ It is easily 
calculated that $\int_{0}^{\tau }\left\langle \psi (t)\right\vert \mathcal{H}%
(t)\left\vert \psi (t)\right\rangle dt=\pm \varpi \delta \cos 2\theta _{0}$,
where the sign + (-) corresponds to the initial state being $\left\vert \uparrow 
\right\rangle
(\left\vert \downarrow\right\rangle).$ The A-A phases here are $\beta _{\uparrow }=-\varpi \delta
(1-\cos 2\theta _{0})$ and $\beta _{\downarrow }=\varpi \delta (1-\cos
2\theta _{0}). $ At time $\tau ,$ the system evolves as 
\[
\left\vert \psi (\tau )\right\rangle =e^{-i\varpi \delta }\exp (-i\theta
_{0}\sigma _{x})\left\vert \uparrow \right\rangle 
\]%
if it is initially in the state $\left\vert \uparrow \right\rangle .$ Likewise, $\left\vert
\psi (\tau )\right\rangle =e^{i\varpi \delta }\exp (-i\theta _{0}\sigma
_{x})\left\vert \downarrow \right\rangle$ for the 
initial state $\left\vert \downarrow \right\rangle$.

\emph{Adiabatic cases.---}Now consider a slowly changing effective
Hamiltonian $\mathcal{H}(t)$, whose instantaneous eigenequation is $\mathcal{%
H}(t)\left\vert n(t)\right\rangle _{e}=\mathcal{E}_{n}(t)\left\vert
n(t)\right\rangle _{e}$. For non-degenerate systems and a 
periodic Hamiltonian $%
\mathcal{H}(\tau )=\mathcal{H}(0)$, the adiabatic theorem shows that 
\[
\left\vert \psi (\tau )\right\rangle _{e}=\exp (i\phi _{n})\left\vert
n(0)\right\rangle _{e} 
\]%
where $\phi _{n}=-\int_{0}^{\tau }\mathcal{E}_{n}(t)dt+\int_{0}^{\tau } $ $%
_{e}\left\langle n(t)\right\vert d\left\vert n(t)\right\rangle _{e}$ for a 
system that is initially in the state $\left\vert n(0)\right\rangle _{e}$. 
For the bare system we have $\left\vert \psi (\tau )\right\rangle =\exp (i\phi )\mathcal{G}%
\left\vert n(0)\right\rangle _{e}$. In this case, both the dressed $\mathcal{%
H}(t)$ and bare $H(t)$ Hamiltonians are periodic.

Now let us consider an ion with two ground states $\left\vert 0\right\rangle
,\left\vert 1\right\rangle $ and one excited state $\left\vert
e\right\rangle $ \cite{Duan}. The Hamiltonian for the ion-laser interaction can be approximated by%
\[
H(t)=\left\vert e\right\rangle (\Omega _{0}\left\langle 0\right\vert +\Omega
_{1}\left\langle 1\right\vert )+\text{h.c.} 
\]%
in the rotating frame, where $\Omega _{0},\Omega _{1}$ are controllable
slow-varying Rabi frequencies. If we chose the operator $\mathcal{G}=\left\vert
1\right\rangle \left\langle 0\right\vert +\left\vert 0\right\rangle
\left\langle 1\right\vert +\left\vert e\right\rangle \left\langle
e\right\vert,$ and if we choose $\Omega _{0}=\cos \frac{\theta }{2}$ $\ $and $%
\Omega _{0}=-\sin \frac{\theta }{2}e^{i\varphi }$ and $\Omega _{1}=\cos 
\frac{\theta }{2}$, then the effective Hamiltonian $\mathcal{H}(t)$ will 
have a dark
eigenstate (state with the zero-eigenvalue) 
\[
\left\vert D\right\rangle =\cos \frac{\theta }{2}\left\vert 0\right\rangle
+\sin \frac{\theta }{2}e^{i\varphi }\left\vert 1\right\rangle. 
\]%
The dynamical phase vanishes in this case while the Berry phase is given by $\Phi =\int $ $\sin \theta d\theta d\varphi $. If the parameters $(\theta ,\varphi )$ undergo a cyclic
evolution, starting and ending at the point $\theta =0$ in the bare
system, then the corresponding evolution will be $\left\vert \psi (\tau )\right\rangle =\exp (i\Phi )\mathcal{G}\left\vert
0\right\rangle$.
In this case the system experiences an evolution from $\left\vert 0\right\rangle $
to $\left\vert 1\right\rangle $, along with an additional all-geometric phase factor, when the dressed system makes a cyclic evolution.

\emph{Superposition of Eigenstates.--- }Since the set of eigenvectors
$\left\{ \Psi _{k}(0)\right\} _{\tau }$ is complete we can 
expand any initial wave function $\Psi(0)$ as 
$\Psi(0) = \sum_k \alpha_k \Psi _{k}(0)$. Despite the fact that 
$\Psi(0)$ is generally not an eigenstate of $W(\tau)$, i.e., $W(\tau)\Psi(0) = \sum_k \alpha_k \exp{(i\phi_k)} \Psi _{k}(0)$, an arbitrary Hamiltonian can 
still be used to execute any given gate $\mathcal{G}$ up to a phase 
$\phi$ at time $\tau$ provided the eigenvalues associated with the states 
$\left\{ \Psi _{k}(0)\right\} _{\tau }$ obey certain conditions. To establish 
these conditions notice that if we require
\begin{equation}
\label{super}
U(\tau)\Psi(0) = \exp{(i\phi)}\mathcal{G}\Psi(0) 
\end{equation}
then we must have
\begin{equation}
\sum_k \alpha_k [\exp{(i\phi_k)} - \exp{(i\phi)} ]\Psi _{k}(0) = 0. \nonumber
\end{equation}
For indices $k$ such that $\alpha_k \neq 0$ there is set of 
requirements imposed on the corresponding eigenvalues, namely 
$\phi = \phi_k + 2\pi m$ for some integer $m$. 

As an example, let us reexamine the case above for 
$\mathcal{G}=\exp (-i\theta _{0}\sigma _{x})$. If we again 
choose the time $t=\tau$ such that $\omega (\tau -\delta )=\theta _{0}$ 
we obtain the two eigenvalues $\exp{(\pm i \varpi \delta)}$ of $W(\tau)$. 
Suppose we expand an arbitrary qubit state $\Psi(0)$ in the basis 
$\Psi _{\uparrow ,\downarrow }(0)=$ $%
\left\vert \uparrow ,\downarrow \right\rangle$ so that 
$\Psi(0) = \cos{(\xi)}\left\vert \uparrow \right\rangle + 
\sin{(\xi)}e^{i \gamma}\left\vert \downarrow \right\rangle$. In 
order to satisfy Eq.~(\ref{super}) the phase $\phi$ must satisfy both 
$\phi = -\varpi \delta +2\pi m$ and $\phi = \varpi \delta +2\pi m^{\prime}$. 
Since this requires that $\varpi \delta = \pi n$ for some integer $n$, the 
evolution operator becomes 
$U(\tau)= \pm \exp{(-i\theta_0 \sigma_x)} = \pm \mathcal{G}.$

\begin{figure}[thp]
\includegraphics[width=.45\textwidth]{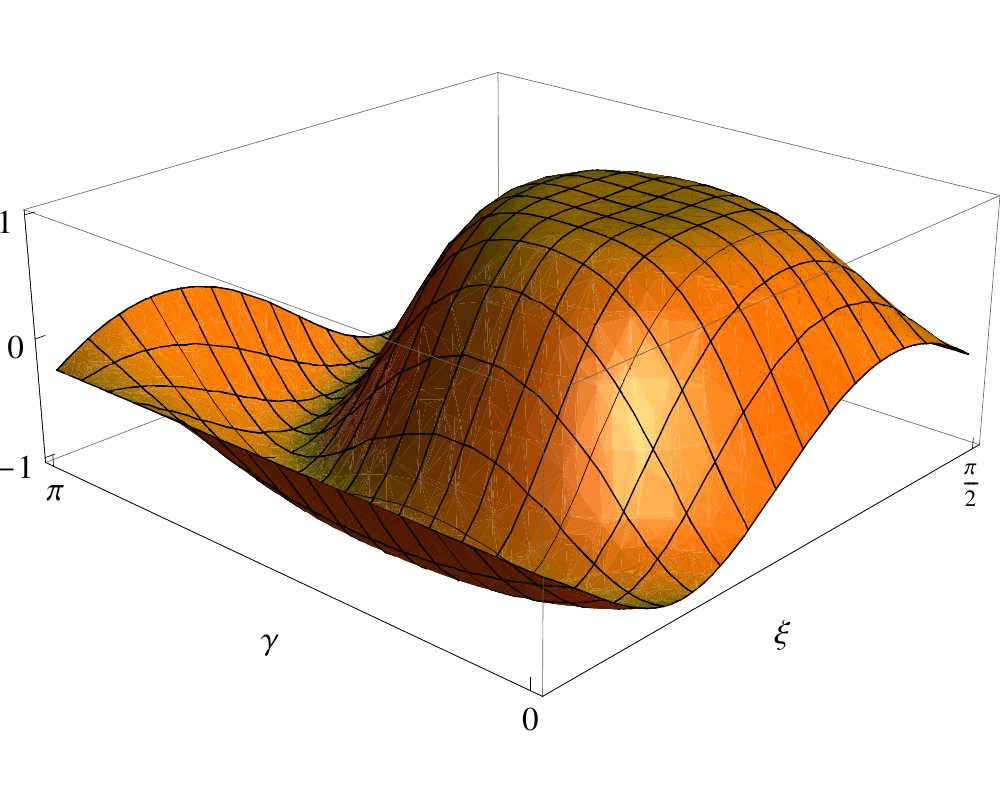}
\caption{The real part of the dressed phase $e^{i\beta}$ 
acquired by the initial state $\Psi(0) = \cos{(\xi)}\left\vert \uparrow 
\right\rangle + 
\sin{(\xi)}e^{i \gamma}\left\vert \downarrow \right\rangle$ after 
evolving according to Eq.~(\ref{super}). In this example, we illustrate the behavior 
of the dressed phase for the gate $\mathcal{G}=\exp (-i\theta _{0}\sigma _{x})$}.
\label{fig:phase}
\end{figure}

For an arbitrary initial state $\Psi(0)\rightarrow \Psi(t)$ we have  
$\int_{0}^{\tau }\left\langle \Psi (t)\right\vert \mathcal{H}%
(t)\left\vert \Psi (t)\right\rangle dt= \theta_0 \sin{(2\xi)}\cos{(\gamma)} 
+ \delta\varpi [\cos{(2\xi)}\cos{(2\theta_0)} + \sin{(2\xi)}\sin{(2\theta_0)}
\sin{(\gamma)}]$. Since $\varpi \delta = \pi n$ and $\phi =\pi n $ for 
some integer $n$, the geometric phase angle acquired during the subsequent 
evolution of $\Psi(0)$ is given by $\beta = \pi n + \theta_0 \sin{(2\xi)}\cos{(\gamma)} 
+ \pi n [\cos{(2\xi)}\cos{(2\theta_0)} + \sin{(2\xi)}\sin{(2\theta_0)}
\sin{(\gamma)}]$. Fig.~\ref{fig:phase} shows the real part of the 
geometric phase $e^{i\beta}$ as a function of the initial 
state parameters $\xi$ and $\gamma$ for $n=\theta_0=1$. 
 
Having a knowledge of the geometrical phase acquired by an arbitrary 
quantum state as 
it undergoes a unitary evolution can help one to predict the outcome 
of an interference experiment designed to test the effects 
associated with the interaction of two or more systems. We will discuss 
the interference effects which should be obtained in an experiment 
involving two bosonic spin chains next. Although we limit our discussion 
to only two chains here, the generalization to an arbitrary 
number of chains should be a straighforward.

\emph{Experimental demonstration.--- }It was recently shown that interference can arise in a two-dimensional 
bosonic lattice when a quantum state is transferred perfectly from 
one site to another \cite{wupst}. An experimental setup based on 
this result can be used to
demonstrate the geometric phase according to the field intensity at 
an appropriate 
site. 
\begin{figure}[thp]
\includegraphics[width=.25\textwidth]{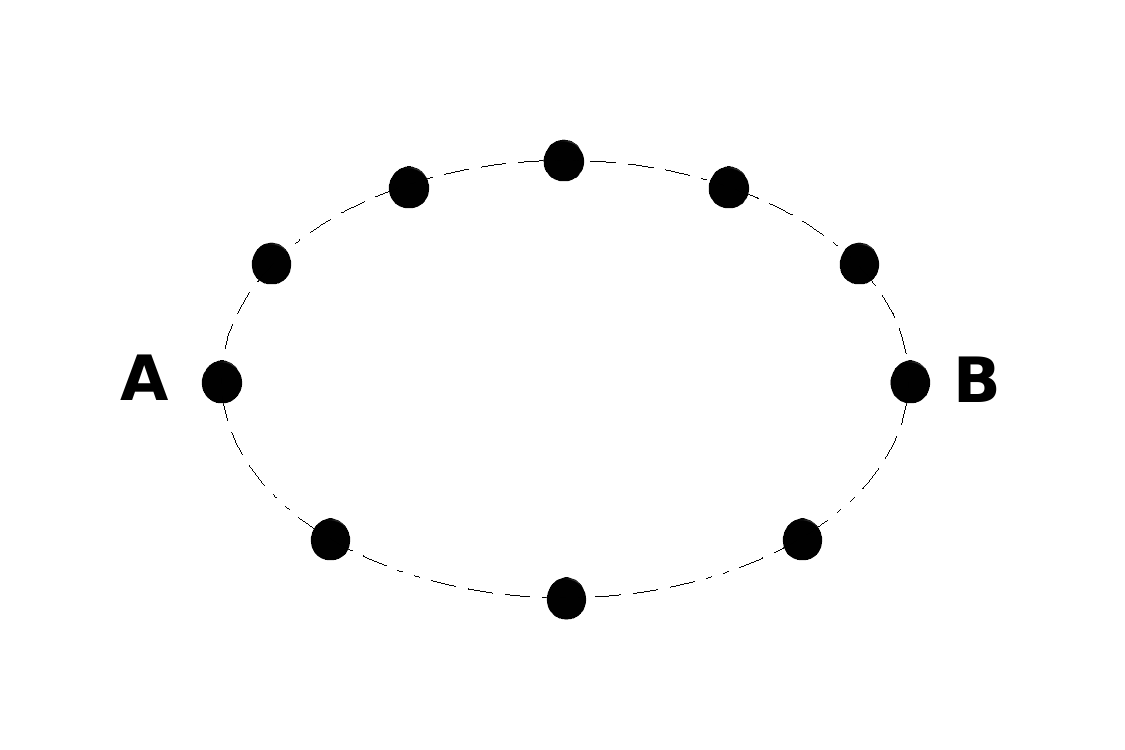}
\caption{Schematic of a boson ring which can be used to demonstrate the geometric phase according to the intensity at site B. Quantum interference 
can arise at site B since the optical 
path lengths of the upper and lower paths are equal but contain 
different numbers of atoms.}
\label{fig:ring}
\end{figure}
To exemplify this effect imagine a ring of bosonic atoms as depicted in 
Fig.~\ref{fig:ring}. The ring can be thought of as two separate 
chains which share first and last sites, denoted respectively  
by A and B in the figure. Both the upper and lower chains are 
assumed to have equal length but the number of sites 
contained in either chain can be different. In our example here 
we have $N_U=7$ and $N_L=5$ sites for the upper and lower paths. 
We will consider the dynamics of bosons governed by the Bose-Hubbard 
Hamiltonian
\[
H= -\sum_{<i,j>} J_{i,j} ({\hat{a}}_i^{\dagger}{\hat{a}}_j +
{\hat{a}}_j^{\dagger}{\hat{a}}_i) +U\sum_i {\hat{n}}_i({\hat{n}}_i-1) + 
\sum_i \epsilon_i{\hat{n}}_i,
\]
where $<i,j>$ indicates that the sum is restricted to nearest 
neighbors in the lattice and ${\hat{a}}_i^{\dagger} ({\hat{a}}_i)$ 
denotes the creation (annihilation) operator of a boson at 
site $i$. Also, ${\hat{n}}_i = {\hat{a}}_i^{\dagger}{\hat{a}}_i$ gives the 
total number of bosonic atoms at site $i$. This Hamiltonian 
can allow for a perfect state transfer through both 
the upper and lower paths of the ring (see 
Ref.~\cite{wupst}). When this occurs at time $t_0$ we have
\[
U^{\dagger}(t_0){\hat{a}}_{i}^{\dagger}U(t_0) = r{\hat{a}}_{N-i+1}^{\dagger},
\]
where $r=\exp(-i\pi(N_U-1)/2)$ for the upper path and 
$r=\exp(-i\pi(N_L-1)/2)$ for the lower path. The expression above is given 
for a linear chain containing $N$ sites. It should be understood that 
in the situation we are considering, 
where the chain is not open-ended but instead forms a closed loop, two 
indices $i$ and $i^{\prime}$ should be used above, one for the upper 
chain and one for the lower. Now, let us expand the 
field operators at $t=0$ in the Wannier basis
\[
\psi({\bf{x}}) = \sum_i [{\hat{a}}_{i}^U w({\bf{x}}-{\bf{x}}_{i}^U)+ {\hat{a}}_{i}^L w({\bf{x}}-{\bf{x}}_{i}^L)],
\]
where ${\hat{a}}_{i}^U$ and ${\hat{a}}_{i}^L$ act on the upper and lower chain, respectively. We see that the operators will evolve to 
\[
\psi({\bf{x}},t_0) = \sum_i [r{\hat{a}}_{i}^U w({\bf{x}}-{\bf{x}}_{N_U -i+1}^{U}) + {\hat{a}}_{i}^L w({\bf{x}}-{\bf{x}}_{N_L-i+1}^L)]
\]
at time $t_0$. Here we have set $r=1$ for the lower path. Although $r=-1$ for the upper path in Fig.~\ref{fig:ring}, we are keeping the expression in a 
more general form to in order to examine the interference effects when 
the number of sites is varied. The average field intensity at ${\bf{x}}$ is 
given by $I({\bf{x}},t_0) = <\psi({\bf{x}},t_0)^{\dagger}\psi({\bf{x}},t_0)>$, where $<...>$ denotes 
the expectation value for the initial state. This can be calculated to be 
\[
I({\bf{x}},t_0) =<{\hat{a}}_1^{\dagger}{\hat{a}}_1>|w({\bf{x}}-{\bf{x}}_N)|^2(2+r+r^*),
\]
where ${\hat{a}}_{1}^U = {\hat{a}}_{1}^L \equiv {\hat{a}}_1$ and ${\bf{x}}_{N_U}^U= {\bf{x}}_{N_L}^L \equiv {\bf{x}}_N$. 

The intensity varies with different values of the signature $r$, in the situation depicted in Fig.~\ref{fig:ring} we have $I({\bf{x}},t_0) = 0$. Constructive, 
destructive, and in-between interference effects can be obtained 
by varying the number of sites in either path.

\emph{Conclusion.--- } We have shown that {\it{any}} Hamiltonian 
can be used to administer {\it{any}} given gate up to a phase during an 
arbitrary time interval provided the system is initialized to an appropriate 
state. The ``dressed'' phase accompanying this evolution is found 
to contain both geometrical and dynamical contributions. The 
evolution of a quantum system determined by the chosen Hamiltonian will not 
necessarily 
be cyclic, nevertheless, we are able to 
determine the geometrical part of this phase using the dressed 
Hamiltonian associated with the gating operation.

\section{Acknowledgments}

L.A.W. was supported by a Ikerbasque Foundation Start-up, the Basque
Government (grant IT472-10) and the Spanish MEC (Project No.
FIS2009-12773-C02-02).  
Part of this material is based upon work supported by NSF-Grant No. 0545798 to
M.S.B.


\begin{thebibliography}{28}




\bibitem{Bose} S. Bose, Phys. Rev. Lett. \textbf{91}, 207901
(2003).

\bibitem{AA} Y. Aharonov and J. Anandan, Phys. Rev. Lett. \textbf{58}, 1593
(1987).

\bibitem{Wilcek} A. Shapere and F. Wilcek, {\it{Geometric Phases in Physics}}, 
Advanced Series in Mathematical Physics Vol. 5 (World Scientific, Singapore, 
1989).




\bibitem{Berry} M.V. Berry, Proc. R. Soc. London Ser. A \textbf{392}, 45 
(1984).

\bibitem{Wu94} L.-A. Wu, Phys. Rev. A \textbf{50}, 5317 (1994).

\bibitem{Christandl} M. Christandl, N. Datta, A. Ekert, and A. J. Landahl,
Phys. Rev. Lett. \textbf{92}, 187902 (2004).

\bibitem{wupst} L.-A. Wu, A. Miranowicz, X. B. Wang, Y. X. Liu, and F. Nori,
Phys. Rev. A \textbf{80}, 012332 (2009).


\bibitem{Duan} L.-M. Duan, J. I. Cirac, P. Zoller, Science \textbf{292},
1695 (2001).
\end{thebibliography}
\end{document}